Considerations for utilizing sodium chloride in epitaxial molybdenum disulfide

Kehao Zhang[1,2], Brian M. Bersch[1], Fu Zhang[1,2], Natalie C. Briggs[1], Shruti Subramanian[1], Ke Xu[3], Mikhail Chubarov[4], Ke Wang[5], Jordan Lerach[5], Joan M. Redwing[1,4], Susan K. Fullerton-Shirey[3,6], Mauricio Terrones[1,2,7], Joshua A. Robinson[1,2,4] *

1. Department of Materials Science and Engineering and Center for Two Dimensional and Layered Materials, The Pennsylvania State University, University Park, PA 16802 USA
2. Center for Atomically Thin Multifunctional Coatings, The Pennsylvania State University, University Park, PA 16802 USA
3. Department of Chemical and Petroleum Engineering, University of Pittsburgh, Pittsburgh, PA 15213 USA
4. 2-Dimensional Crystal Consortium (2DCC), Materials Research Institute, The Pennsylvania State University, University Park, PA 16802, USA
5. Materials Characterization Laboratory, The Pennsylvania State University, University Park, PA 16802 USA
6. Department of Electrical and Computer Engineering, University of Pittsburgh, Pittsburgh, PA 15213 USA
7. Department of Physics, Chemistry, The Pennsylvania State University, University Park, PA 16802 USA
    * jrobinson@psu.edu

## Abstract

The utilization of alkali salts, such as NaCl and KI, have enabled the successful growth of large single domain and fully coalesced polycrystalline two-dimensional (2D) transition metal dichalcogenide layers. However, the impact of alkali salts on photonic and electronic properties are not fully established. In this work, we report alkali-free epitaxy of $MoS_2$ on sapphire and benchmark the properties against alkali-assisted growth of $MoS_2$. This study demonstrates that although NaCl can dramatically increase the domain size of monolayer $MoS_2$ by 20 times, it can also induce strong optical and electronic heterogeneities in as-grown large-scale films. This work elucidates that utilization of NaCl can lead to variation in growth rates, loss of epitaxy, and a high density of nanoscale $MoS_2$ particles ($4 \pm 0.7/\mu m^2$). Such phenomena suggest that alkali atoms play an important role in Mo and S adatom mobility and strongly influence the 2D/sapphire interface during growth. Compared to alkali-free synthesis under the same growth conditions, $MoS_2$ growth assisted by NaCl results in >1% tensile strain in as-grown domains, which reduces photoluminescence by ~20× and degrades transistor performance.

## Introduction

Two-dimensional (2D) transition metal dichalcogenides (TMDs)[1,2] are primed for a wide variety of promising technologies in nanoelectronics (field effect transistors[3,4], phototransistors[5]), photonics (photo-detectors[6], light emitting diodes[7]) and sensors (gas sensors[8], bio-sensors[9]). However, most devices are based on exfoliated flakes, which are not compatible with industrial needs. Therefore, many efforts continue to focus on the synthesis of large scale TMDs.[2,10–14] Ion-exchange methods enabled wafer scale production of molybdenum disulfide ($MoS_2$)[15] and tungsten diselenide ($WSe_2$)[16], but these films suffer from poor electrical performance due to a nanocrystalline structure (<100 nm domain size). On the other hand, vapor phase deposition with nucleation seeds such as $MoO_3$ nanoribbons and perylene-3,4,9,10-tetracarboxylic acid tetrapotassium salt (PTAS) has enabled >100 μm single crystal domain size.[17–19]

Recent utilization of alkali salts clearly demonstrates that its use enables 5~100× increases in domain size[12,13] and is reported to be the key to realizing the first 100mm wafer synthesis of polycrystalline monolayer $MoS_2$.[11] Most recently, using NaCl helped exploring the novel 2D materials.[20] The addition of NaCl is said to increase the single crystalline domain size of monolayer $MoS_2$ in metal organic chemical vapor deposition (MOCVD) due to dehydration[12] and nucleation suppression.[13] For traditional solid-source (e.g. $MoO_3$ and Mo foil) CVD vaporization, NaCl is demonstrated to increase the reactivity of metal and chalcogen elements, which is the key to enable the batch production of 6-inch uniform monolayer $MoS_2$[11] and the exploration of an entire TMD library,[20] where in some cases, large area 2D films cannot be achieved without NaCl. Although >100 μm single crystalline TMDs are readily synthesized via alkali-assisted methods, it is highly unlikely that nucleation can be suppressed to a single nuclei on a substrate that subsequently grows to multiple centimeters. Therefore, large scale epitaxy has recently been the focus to realize large scale TMDs films with minimum grain boundaries.[14,21,22] Building from this, it may be hypothesized that using alkali-assisted epitaxy may be a route to achieving epitaxial TMD films with large domain size. However, alkali metals are also proven impurities that must be avoided at all cost in traditional silicon-based technologies due to high rates of ion diffusion through the gate oxides, which leads to unreliable performance.[23–29] While the utilization of alkali-assisted growth continues to expand among the community,[11–13,19,20] a direct comparison of the impact of alkali metals on the electronic and photonic properties $MoS_2$ films (with and without layer transfer) has not been achieved, therefore a comprehensive study on the impact is merited.

In this work, we explore the trade-offs of utilizing NaCl in MOCVD synthesis of coalesced, epitaxial $MoS_2$ monolayer films. We directly benchmark the optical, structural and electronic properties of alkali-free and alkali-assisted $MoS_2$ grown under the same conditions. Monolayer films grown *without* NaCl are found to be epitaxial and uniform across a 2×2 $cm^2$ sapphire substrate, while NaCl-assisted growth is shown to enhance the growth rate and domain size by 20× for some domains. Meanwhile, residual strain in NaCl-assisted films leads to a 20× quench in photoluminescence (PL) of the large domains. Additionally, when using the same growth parameters as alkali-free MOCVD, we find a high density of multi-layer $MoS_2$ particles and an atomic layer of Na at the $MoS_2$/sapphire interface that destroys the epitaxial relationship between $MoS_2$ and sapphire – precluding the ability to achieve large-scale, epitaxial 2D layers. Moreover, we compare transport properties films transferred to $SiO_2$/Si substrates to understand if the variations found in as-grown films translates to films removed from the substrate. Utilizing NaCl introduces a 1.5× decrease in mobility, 2× increase in subthreshold slope, and 100× reduction in on/off ratio when directly compared to alkali-free growths. We also find a 3× higher variation (>15V) in threshold voltage in transistors that utilize alkali-assisted $MoS_2$, suggesting that any residual Na remaining from the transfer process dominates the transistor performance.

**Results and Discussion**

Optimization of alkali-free metal organic chemical vapor deposition (MOCVD; Fig. S1) on c-plane (0001) sapphire enables the synthesis of highly uniformity, stoichiometric, epitaxial $MoS_2$. To achieve full coalescence, films are synthesized using 1.2×10⁻³ sccm of molybdenum hexacarbonyl (Mo(CO)$_6$) and 2.71 sccm diethyl sulfide (DES) with 565 sccm Ar as carrier gas at 900°C, 10 Torr for 2 hours. Figure 1a presents a schematic of the MOCVD reactor, and photo of a typical sample (Fig. 1b) demonstrating uniform coverage across a 2×2 cm². Low magnification scanning electron microscopy (SEM) (Figure 1c) confirms the uniformity of the film, with atomic force microscopy (AFM) characterization confirming the film is particle free with < 0.5nm surface roughness (Figure 1d). Note that secondary MoS$_2$ layers are also observed (white arrow), and the height variation in the AFM image is due to sapphire terraces.[30] X-ray photoluminescence spectroscopy (XPS) shows Mo-S bonding at 229.7 eV and 232.8 eV for Mo 3d and 162.4 eV and 163.7 eV for S 2p, and shows no Mo-O or S-O bonding often found in powder based synthesis (Fig. 1e).[31] Quantitative analysis reveals the S/Mo ratio is 1.95 (+/- 0.1), with a low density of S vacancies (Fig S3). Importantly, in-plane x-ray diffraction (XRD) verifies the epitaxial nature of the MoS$_2$ (Fig. 1f), as indicated by the presence of 60° separated peaks in rotation angle around the sample surface (φ) for a diffraction angle (2θ) set to (10$\bar{1}$0) MoS$_2$ (Fig. 1f). This observation, together with the measurement of the φ scan, for {11$\bar{2}$0} α-Al$_2$O$_3$ peaks demonstrate that MoS$_2$ is 30° rotated compared to the Al$_2$O$_3$ substrate, with an in-plane epitaxial relation being (10$\bar{1}$0) α-Al$_2$O$_3$∥(10$\bar{1}$0) MoS$_2$. To benchmark alkali-assisted and alkali-free MOCVD of MoS$_2$, we utilize these same growth conditions, and place 5 mg NaCl powder 5 cm upstream of the substrate to introduce NaCl vapor into the chamber during growth.

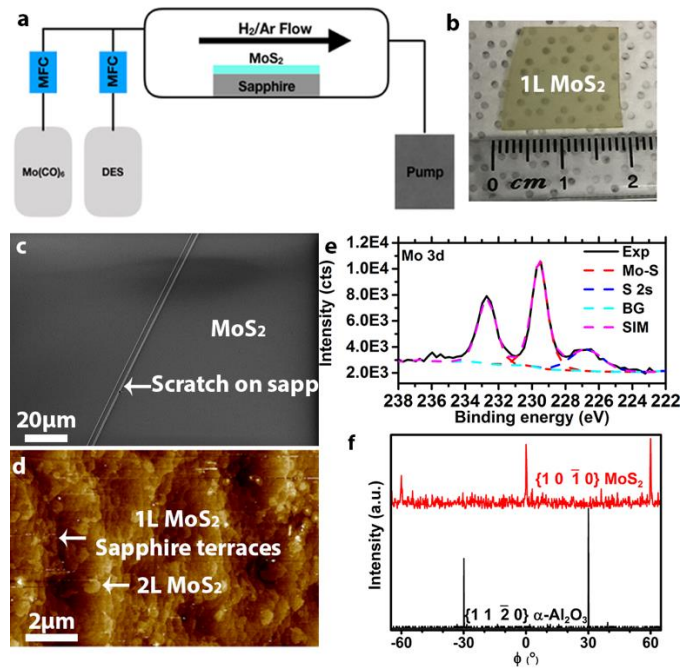

**Figure 1-Large area epitaxy of MoS$_2$.** a. Schematic of the MOCVD reactor, the substrate is loaded in a hot wall tube reactor with Mo(CO)$_6$ and DES as precursor; b. photograph of monolayer MoS$_2$ grown on a 2cm² sapphire substrate; c-d SEM (c) and AFM (d) images of 1L MoS$_2$, confirming the uniformity and low roughness; e. XPS spectra of synthetic MoS$_2$ in Mo 3d range showing the Mo-O bonding is below the detection limit (<0.5%); f. In-plane XRD φ scans of {11$\bar{2}$0} α-Al$_2$O$_3$ bottom lower curve and {10$\bar{1}$0} MoS$_2$ upper red curve

The addition of NaCl can lead to heterogeneous growth kinetics, evident by comparing MoS$_2$ layers grown with and without NaCl as a function of time and NaCl temperature. When NaCl is used, under *all* circumstances, epitaxy is lost between the MoS$_2$ and sapphire. When NaCl is kept at 500°C or below, there is no significant increase in domain size. Once NaCl is heated to 600°C, similar to previous reports,[13] there is a distinct >20× increase in domain size compared to the growth without NaCl (Fig. 2a and Fig. S4, S5); however, this trend does not extend beyond very short grow times. SEM images of $MoS_2$ after a 2 min growth demonstrates that domains grown with NaCl are >1 μm in lateral size (Fig. 2a) compared to ~50 nm for alkali-free (Fig. 2d). However, when the growths are extended beyond 2 min, we find the large $MoS_2$ domains (L-$MoS_2$ hereafter) do not continue growing. Instead small $MoS_2$ domains (S-$MoS_2$ hereafter) nucleate and grow to fill the gaps between L-$MoS_2$ to form a coalesced monolayer film after 60 min (Fig. 2b, c). In comparison, the alkali-free $MoS_2$ uniformly nucleates across the surface as S-$MoS_2$ domains (Fig. 2e) that continue to grow at an even pace to form epitaxially oriented film at 60 mins (Fig. 2f). We note that secondary $MoS_2$ layers also nucleate and grow regardless of growth conditions, appearing as dark spots in the SEM images. To identify if NaCl vapor pressure impacts the growth, we modulate the NaCl vapor concentration during the growth by tuning the temperature of NaCl. We find that under all conditions tested (low to high NaCl vapor concentration: $2.52\times10^{-5}$~140sccm)), there is a comparable L-$MoS_2$ and S-$MoS_2$ distribution for the alkali-assisted growths (Figure S4, S5), indicating that the NaCl impacts on adatom mobility (film morphology) is similar at a wide range of NaCl vapor concentration. Comparing Raman spectra of L- and S-$MoS_2$, indicates that the L-$MoS_2$ domains exhibit a uniform 1.5 cm$^{-1}$ red shift of the $E_{2g}$ peak (Fig. 2g) compared to S-$MoS_2$, indicating the L-$MoS_2$ exhibit ~1.5% higher tensile strain. [37] The peak variation is evident in Raman spectra extracted from five points across the film in Figure 2g (sp1-sp5) and plotted in Figure 2h. The $E_{2g}$ peak position is located at 384.5 cm$^{-1}$, 383.9 cm$^{-1}$ and 384.6 cm$^{-1}$ on spot 2, 3 and 4, respectively, while it is 385.9 cm$^{-1}$ on spot 1 and 5. Furthermore, the PL from the interior/edge region of L-$MoS_2$ domain (Figure 2i,j) is 5-20× lower in intensity. This PL quench may be partially due to the observed strain as suggested by Raman (Fig.2h), however, it cannot be solely due to strain-induced band structure change since there is no peak shift in the PL.[32,33] One possible scenario can be strain modifies the radiative recombination rate, which is also reported in low-dimensional materials.[34] In contrast, the alkali-free growth exhibits a laterally uniform PL response (Figure S6), suggesting that the presence of the NaCl during growth can lead to modified interfacial interactions that lead to mechanically and optically non-uniform $MoS_2$ films. This indicates that uniform films consisting of L-$MoS_2$ are likely to exhibit uniformly degraded PL across the surface compared to an alkali-free grown film because of the modified 2D/sapphire interface.

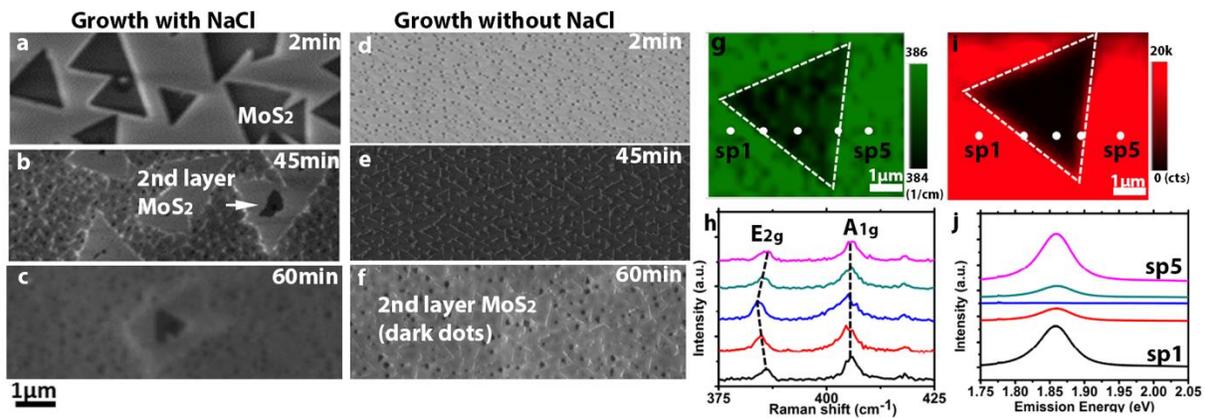

**Figure 2-Growth kinetics heterogeneity of $MoS_2$ grown with NaCl.** a-f. SEM images of the $MoS_2$ monolayer grown with (a-c) and without NaCl (d-f), respectively; g. $E_{2g}$ position map of $MoS_2$ film, clear red shift in a ~3 μm triangular region is observed, indicating stronger tensile strain in this domain; h. representative Raman spectra taken

from the labeled spots in the map. All the spectra show the signature peaks of $MoS_2$, confirming the existence of $MoS_2$, and $E_{2g}$ peak is 384.5cm$^{-1}$, 383.9cm$^{-1}$ and 384.6cm$^{-1}$ on spot 2, 3 and 4 respectively, while it is 385.9 cm$^{-1}$ on spot 1 and 5; i. PL intensity map of the $MoS_2$ film, the PL is significantly quenched in the triangular region due to the strain induced band structure change; j. representative PL spectra taken from labeled spots. The PL in the triangular domain is 20× quenched along with 5× quench at the edge of the large domain.

Another distinct difference between alkali and alkali-free growth is the resulting surface morphology. Scanning electron micrographs (Fig 2) suggest that for fully coalesced monolayer films, "islands" form on the surface of NaCl-assisted films. AFM (Fig. 3a,b) confirms a smooth, uniform growth of $MoS_2$ without NaCl, indicating in-plane growth is preferred throughout the synthesis time. However, NaCl-assisted films contain a mixture of 2D layers and nanometer-thick particulates (particulates density ~4±0.7 /μm$^2$), suggesting a convolution of growth modes (island and layer-by-layer) in the presence of NaCl vapor. Based on TEM and EDS, the particles are found to be multi-layered $MoS_2$ (Fig. 3c-e), suggesting that NaCl can lead to a wide range of film morphologies based on synthesis parameters. We note that while others have not directly discussed the presence of particles, they are readily observed for both alkali-assisted solid-source CVD and MOCVD.[13,20] This suggests that, regardless of growth conditions, precursor or substrate, the presence of alkali-ions leads to heterogeneous growth.

Regardless of growth parameters, the deposition of alkali-metal atoms onto the substrate surface is unavoidable. Similar to Kim *et al.*,[13] XPS analysis of the $MoS_2$ grown with NaCl clearly shows a strong Na 1s, with the Cl ls peak below the detection limit (Fig. S7). Under all conditions examined, the Na 1s peak at 1072.3 eV, indicates the majority of the Na bonding is associated with Na-O,[35] with the Na-S interaction is below the XPS detection limit, suggesting that Na is deposited first on the sapphire substrate and acts as a catalyst for growth similar to prior works.[11,13] However, we additionally identify that the Na is not distributed uniformly across the substrate surface. Using time-of-flight secondary ion mass spectroscopy (TOF-SIMS) mapping and atomic-resolution depth profiling (Fig. 3f), it is evident that the location of the Na varies across all axes (x,y, and z). Similar to the prior work using TOF-SIMS to reveal the interfacial elemental information on CVD graphene,[36] Monitoring the mass intensity signal for Na, Mo, and Al, the relative concentration of Na, $MoS_2$, and sapphire can be mapped as a function of depth. Utilizing TOS-SIMS depth profile maps (Fig. S8), we find that the Mo/Al intensity expectedly decreases/increases uniformly and monotonically with each sputtering cycle, where 1 sputter cycle removes > 80% (based on intensity change) of the $MoS_2$ film. Interestingly, after the first cycle, the Na signal (Fig. 3f, red curve) increases non-uniformly across the surface, with a greater increase in S-$MoS_2$ regions compared to L-$MoS_2$ (Fig. S8). Figure 3f (inset) is a Na signal intensity map, showing wide-spread, non-uniform coverage of Na-ions on the surface following removal of the $MoS_2$ layer from the surface. The higher Na concentration beneath S-$MoS_2$ regions (Fig.3f inset, bright yellow area) indicates that a layer of Na is present at the $MoS_2$/sapphire interface in these regions. However, the lack of a strong Na signal below L-$MoS_2$ domains (Fig. 3f inset, brown triangles) indicates a much lower density of Na in these regions. These findings correlate well with AFM data (Fig. 3g) that shows the L-$MoS_2$ has a lower z-height value than its surrounding S-$MoS_2$ counterparts, indicating the surrounding S-$MoS_2$ is higher than L-$MoS_2$ – likely from a Na-interlayer.

Combining AFM, TEM, XPS, and TOF-SIMS (Fig. 3, Fig. S7, Fig. S8) leads to the hypothesis that Na plays multiple roles in the nucleation, growth and properties of the $MoS_2$. At the very early stages of the growth, the Na only partially passivates the sapphire surface with low density Na-O bonds (based on XPS

(Fig. S7)) that initially disrupt the epitaxial relationship (based on AFM/SEM, Fig. 2,3), but subsequently enhance growth rate by lowering the energy barrier for domain-edge Mo-S bonding and catalyze stronger interfacial interactions (based on Raman, Fig. 2).[11] However, as the growth time goes beyond the initial stages, the substrate surface becomes fully saturated with Na-O bonds that passivate the in-plane L-$MoS_2$ growth front (also contributed by the enhanced tensile strain) and dramatically reduces the surface energy of the sapphire similar to that of polymer functional layers.[37] With a reduction in surface energy, one would expect a dramatically reduced growth rate, preference for molecular cohesion over substrate adhesion, and the loss of an epitaxial relationship between the film and substrate. This is the case for S-$MoS_2$. Based on TOF-SIMS (Fig.3f), S-$MoS_2$ nucleates and grows *on-top-of* the Na-O layer evident by the 0.2 nm height difference between the L-$MoS_2$ and S-$MoS_2$ (Figure 3g). Because of this, the S-$MoS_2$ exhibits a low growth rate (Fig. 2), high density of multi-layer particles (Fig. 3b), and loss of epitaxy (Fig. 2,3). Figure 3h presents a hypothesized schematic of NaCl-assisted CVD induced growth heterogeneities.

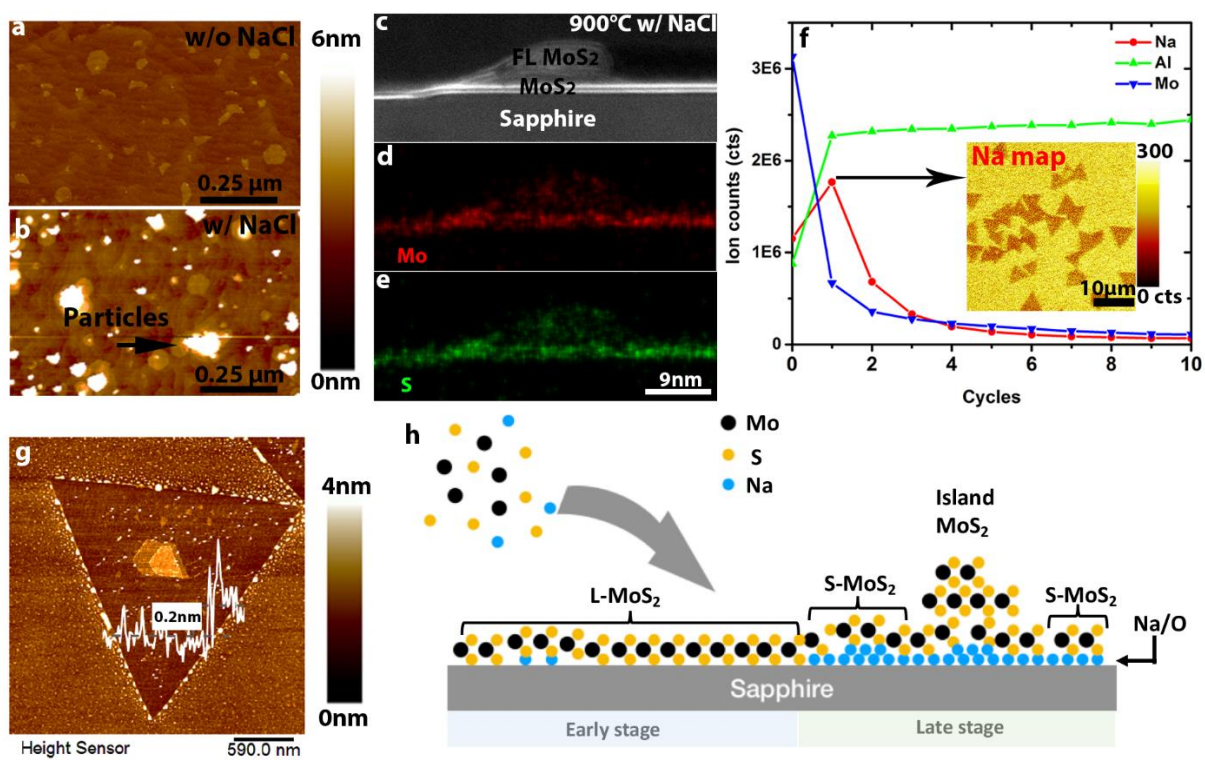

**Figure 3-Growth kinetics and sodium passivation.** a-b. AFM images of the $MoS_2$ monolayer grown with and without NaCl respectively. A high particle density is observed on the film grown with NaCl; c-e. Cross-sectional STEM image focus on the particles on the $MoS_2$ film, revealing that the nature of the particles is multilayer $MoS_2$; f. Elemental depth profile of $MoS_2$ grown with NaCl, the inversely relation between Mo and Na suggest that the Na is underneath the $MoS_2$ (inset: Na elemental map after the 1$^{st}$ cycle of sputtering that removes most of the $MoS_2$); g. AFM image of a large $MoS_2$ domain surrounded by S-$MoS_2$. The inserted height profile suggests the L-$MoS_2$ is 0.2nm thinner than S-$MoS_2$; h. Hypothesized mechanism of NaCl-induced growth heterogeneities.

Non-uniformities in alkali-assisted MOCVD $MoS_2$ films lead to degraded and variable electrical performance. This is true regardless of device type based on evaluation of as-grown films on sapphire with electrolyte-gated (EG) (gate materials: poly (ethylene oxide) and $CsClO_4$) field effect transistors

(FETs) (details about the electrolyte gate is described elsewhere[38]) [30,31,38,39] *and* transferred films on SiO$_2$/Si using back-gated (BG) FETs (See Fig. S9 for transfer details). Figure 4a provides representative transport characteristics of as-grown, electrolyte gated, FETs. Clear differences in threshold voltage and subthreshold swing (SS) between the films grown with (noted with "NaCl" subscript) and without NaCl (no notation) are observed. Detailed FET analysis reveals the field effect mobility is similar in both as-grown cases ($\mu_{NaCl}$ = 3.8±0.8 cm$^2$/V.s; $\mu$ = 3.4 ±1.3 cm$^2$/V.s). However, the NaCl-assisted films exhibit a degraded SS by 2× (SS$_{NaCl}$ = 504±30 mV/dec; SS = 238±6 mV/dec), with a threshold voltage ($V_{th}$) shift of +2 V. The reduction in SS can be expected when considering the heterogeneity of the NaCl-assisted films, however, the $V_{th}$ shift is likely due to the presence of a Na-O interface layer. Typically, synthetic MoS$_2$/c-sapphire exhibits significant charge transfer (electrons transfer from sapphire to MoS$_2$) that heavily n-type dope the MoS$_2$ due to strong film/substrate coupling.[31] In our case, when using NaCl during the growth, the Na-O interface likely suppresses charge transfer from the substrate and thereby decreases the electronic signature of n-type doping.

Prior work claims that transferring the MoS$_2$ from the growth substrate to pristine SiO$_2$/Si substrates eliminates adverse impacts Na may yield.[13] To directly test this theory, we fabricate conventional back-gated FETs from transferred MoS$_2$ on a 300nm SiO$_2$/Si substrate. While the performance of devices can vary widely based on the quality of the transfer technique, relative changes between NaCl-free and NaCl-assisted MoS$_2$ can be readily extracted. In the transferred-film case, we find that in addition to a degraded SS for MoS$_{2(NaCl)}$ (SS$_{NaCl}$ = 11.2±1.5 V/dec; SS = 6.2±1.1 V/dec), the field effect mobility of MoS$_{2(NaCl)}$ is degraded by > 30% compared to MoS$_2$ ($\mu$ =6.0±1.3 cm$^2$/V.s; $\mu_{NaCl}$ =3.8 ±0.7 cm$^2$/V.s) (Fig. 4a,b,e). These mobility values fall in the range of reported room-temperature measurements (0.02~30 cm$^2$/V.s) on single or poly crystalline MoS$_2$.[11–13,17,18,20,31,37,40,41] The most significant impact may be a >100× reduction of on/off ratio (on/off$_{NaCl}$ = 10$^4$~10$^5$; on/off =10$^6$~10$^7$) (Fig. 4b,d), a ~20V shift in $V_{th}$, and dramatic increase in $V_{th}$ variability (Fig. 4c), suggesting that the Na may still impact the interface even after transfer. Similar to as-grown films, the heterogeneity of the film likely plays a role in the wide ranging variability in FET characteristics, however, one would expect the interface heterogeneity to be eliminated upon film transfer if the Na is dissolved in the water during transfer.[13] We find that this is not the case, in fact, the device-to-device variation is magnified compared to as-grown films, suggesting that even if Na is removed, it has altered the intrinsic properties of the MoS$_2$, potentially creating charge trap states within the MoS$_2$ layers themselves.[42–45] To verify the discussed trends, we randomly measure 5 EG-FETs (due to the long charging/reset time) and 9 BG-FETs for each condition and create quartile plots of on/off ratio and mobility (Fig. 4d,e), validating the previous discussion on as-grown and transferred MoS$_{2(NaCl)}$. We note that the differences discussed here cannot be attributed simply to variations between gated measurements because the results are repeatable in both the electrolyte-gated, non-transferred devices and in back-gated, transferred devices, as shown in Fig. S10 and Fig. S11. In the transferred device the interaction between film and substrate is weaker, hence the gate voltage has a stronger effect over the transport in the channel, and we were able to more easily observe the change in mobility and on/off ratio related to the addition of dopants. Unlike devices on exfoliated and transferred TMDs, synthetic TMDs on 3D crystalline substrates always exhibit stronger film/substrate coupling such as chalcogen passivating interlayer[30], charge transfer[31] and Coulomb screening.[46] This unique coupling is a convoluting factor when trying to understand how dopants impact the threshold voltage, mobility and on/off ratio in as grown cases.[31,46]

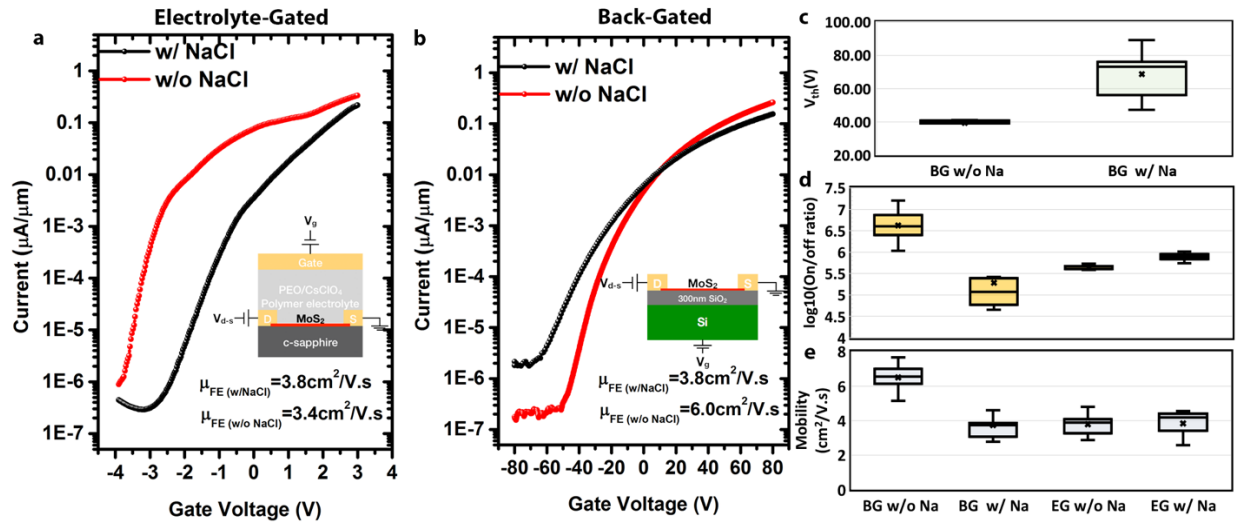

**Figure 4-Heterogeneities in electrical properties of alkali-assisted MoS$_2$.** a. Representative transport curves of electrolyte-gated monolayer MoS$_2$ FETs. A clear positive shift and degradation in subthreshold slope is observed due to the Na intercalation; b. Representative transport curves of back-gated monolayer MoS$_2$ FETs. The on/off ratio is reduced by two orders of magnitude with NaCl as well as 2× degradation in subthreshold swing; c. Quartile plot of threshold voltage of MoS$_2$ FETs; d. Summary of on/off ratio of FETs with as grown and transferred MoS$_2$; e. Summary of mobility of FETs with as grown and transferred MoS$_2$. It is suggested that sapphire substrate can limit the performance in on/off ratio and mobility. Understanding the interface properties is critical to achieve high performance synthetic 2D nano-devices.

In summary, our work highlights a variety of trade-offs that one must consider when utilizing NaCl in monolayer MoS$_2$ epitaxy. Alkali-assisted growth clearly enables a significant increase in the growth rate and domain size by reducing energy barriers for precursor vaporization, enhancement of adatom mobility, and reduction of Mo-O bonding during synthesis. However, this work also presents evidence that use of NaCl destroys the epitaxial relationship between MoS$_2$ and sapphire, and deposits a Na-ion interfacial layer that varies in density based on growth time and NaCl temperature. under slower growing domains, which impacts device performance in as-grown and transferred films. While some heterogeneities in alkali-assisted MOCVD, such as domain growth rates and the density of surface particles, can be overcome by continued optimization of the MOCVD process, evidence suggests that regardless of growth parameters, the loss of epitaxy and Na interfacial layer cannot be avoided. Furthermore, before utilizing alkali metals in the synthesis of 2D semiconductors, one must consider if the end application can tolerate the presence of alkali metals in the required device architectures.

**Acknowledgement**

The work at Penn State was conducted as part of the Center for Atomically Thin Multifunctional Coatings (ATOMIC), sponsored by the National Science Foundation (NSF) division of Industrial, Innovation & Partnership (IIP) under award # 1540018. Partial support of this work comes from DMR-EPM under Grant No. 1607935 and by the Center for Low Energy Systems Technology (LEAST), one of the six SRC STARnet Centers sponsored by MARCO and DARPA.